\documentclass[a4paper,10pt]{article}
\usepackage[a4paper, margin=2cm]{geometry}

\usepackage[utf8]{inputenc}
\usepackage[T1]{fontenc}

\usepackage{fourier}

\usepackage{amsmath}
\allowdisplaybreaks
\usepackage{amsfonts}
\usepackage{bm}
\usepackage{mathrsfs}
\usepackage{pifont}
\usepackage{booktabs}
\usepackage{multirow}
\usepackage{graphicx}
\usepackage{upgreek}
\usepackage{commath}
\usepackage{subfig}

\makeatletter
\newcommand*{\transpose}{%
	{\mathpalette\@transpose{}}%
}
\newcommand*{\@transpose}[2]{%
	\raisebox{\depth}{$\m@th#1\intercal$}%
}
\makeatother

\usepackage{amsthm}

\makeatletter
\def\th@remark{%
	\thm@headfont{\bfseries}%
	\normalfont 
	\thm@preskip\topsep \divide\thm@preskip\tw@
	\thm@postskip\thm@preskip
}
\makeatother

\theoremstyle{remark}
\newtheorem*{remark}{Remark}

\theoremstyle{plain}
\newtheorem{condition}{Condition}

\usepackage[english]{babel}
\usepackage{csquotes}

\usepackage[hidelinks]{hyperref}

\usepackage{color}
\usepackage[normalem]{ulem}

\usepackage[square,numbers]{natbib}
\usepackage{doi}
\usepackage{url}

\title{Shear shock formation in incompressible viscoelastic solids}
\author{H. Berjamin \textsuperscript{a}, S. Chockalingam \textsuperscript{b} \\
{\footnotesize
\begin{tabular}{l}
	~ \\
	\textsuperscript{a}School of Mathematical and Statistical Sciences, NUI Galway, University Road, Galway, Republic of Ireland \\
	\textsuperscript{b}Massachusetts Institute of Technology, Department of Aeronautics and Astronautics, Cambridge, MA, 02139, USA
\end{tabular}}
}
\date{}

\begin{document}


\maketitle

\begin{abstract}
	\noindent
	Experiments have shown that shear waves induced in brain tissue can develop into shock waves, thus providing a possible explanation of deep traumatic brain injuries. Here, we study the formation of shock waves in soft viscoelastic solids subject to an imposed velocity at their boundary. We consider the plane shearing motion of a semi-infinite half-space, which corresponds to a spatially one-dimensional problem. Incompressible soft solids whose behaviour is described by the Fung--Simo quasi-linear viscoelasticy theory (QLV) are considered, where the elastic response is either exponential or polynomial of Mooney--Rivlin--Yeoh type. Waveform breaking can occur at the blow-up of acceleration waves, leading to one sufficient condition for the formation of shocks. A slow scale analysis based on a small amplitude parameter yields a damped Burgers-like equation, thus leading to another shock formation condition. Numerical experiments performed using a dedicated finite volume scheme show that these estimates have limited accuracy. Their validity is restricted in the elastic limit too, where exact shock formation conditions are known. \\
	
	
%
\end{abstract}

\section{Introduction}\label{sec:Intro}

Physical and chemical processes leading to traumatic brain injury (including mild traumatic brain injury, aka concussion) are very complex. In medical terms, primary brain injury results from the transmission of external mechanical forces to the brain \cite{suer20}. Main causes include direct impacts to the skull, rapid acceleration/deceleration of the head, or airborne blast waves.

For all these examples, understanding the effect of external motion on the brain is key to quantify the severity of injuries. In this regard, laboratory experiments performed on gelatin phantoms and brain tissue have shown that smooth acoustic shear waves can develop into shocks in finite time and propagation distance \cite{catheline03,espindola17}. Thus, in specific cases, deep brain injury could be caused by the nonlinear propagation of shear waves in the brain.

In shear mechanical tests, brain tissue exhibits the peculiar property of being very soft \cite{destrade15}. Moreover, it behaves like an incompressible and nonlinear viscoelastic solid (see the review by Budday et al. \cite{budday20} for complements). Up to moderate but finite strains and frequencies, quasi-linear viscoelastic (QLV) solid models are able to reproduce the main features of the mechanical response in an adequate fashion \cite{darvish01, rashid13}. Depending on the supporting literature, such models have been called incompressible Fung or Simo theory \cite{depascalis14,berjamin21}. A two-dimensional variant of Fung--Simo QLV where relaxation of stress is replaced by the relaxation of deformation gradients was also shown reasonably accurate in dynamic configurations with formation of shocks \cite{tripathi19a,tripathi19}.

In incompressible elastic solids, the growth and decay of acceleration waves was presented by Ogden \cite{ogden74}, thus describing the evolution of a wavefront's slope. Connection to the formation of shocks was detailed analytically by Fu and Scott \cite{fu91} in the one-dimensional case, whereas Jordan and Puri \cite{jordan05} performed analytical-numerical comparisons. In the class of incompressible hyperelastic solids, the formation of shocks from plane shear waves was studied by Chu \cite{chu64} using the theory of Riemann invariants, see also Ref.~\cite{chockalingam20} where exact conditions are detailed for various models.

As far as viscoelastic solids are concerned, Coleman and Gurtin \cite{coleman65} investigated the propagation and evolution of acceleration waves in general one-dimensional materials with memory. These results are recalled in the review by Schuler et al. \cite{schuler73}, where subsequent works on the existence of steady shocks are presented (see also Cormack and Hamilton \cite{cormack18} for related results). Lardner \cite{lardner76} studied the formation of shocks in general one-dimensional solids with memory by using an averaging method. Shock wave formation in soft solids with Kelvin--Voigt viscosity was studied by Zabolotskaya et al. \cite{zabo04} based on a slow scale assumption which leads to a modified Burgers equation (note in passing that the viscous stress proposed therein does not yield a frame-invariant model \cite{destrade12}). The latter was then solved numerically using a spectral method with harmonic balance. Lastly, Laksari et al. \cite{laksari14} investigated the evolution of compressional acceleration waves in the context of traumatic brain injury.

In the light of these works, we conduct an analytical-numerical study of plane shear-wave propagation in incompressible QLV solids. Analytical shock formation conditions are based on the study of acceleration waves \cite{ruggeri98} and of reduced wave equations (here, a damped Burgers-like equation, cf. \cite{fusco84,zabo04} for related results). The numerical method is a Godunov-type finite-volume scheme that involves operator splitting, see \cite{tripathi19a,berjamin18} for similar techniques. Analytical-numerical comparisons show that these theoretical estimates are restricted to specific ranges of loading signals. In the elastic limit, comparison with the exact theory \cite{chockalingam20} shows that the slow scale estimates are correct for low Mach numbers. As demonstrated in the recap Table~\ref{tab:Summary}, the limited accuracy of these estimates encourages further research in this direction.

\begin{table}
	\centering
	\caption{Summary. Shock formation estimates derived in this study (equation numbers), where exponential and cubic polynomial solids are considered. Predicted occurrence of shocks at finite propagation times and distances from a boundary subjected to ramp-type or sinusoidal loading (yes, no, or conditional case y/n with both possibilities). Accuracy of these analytical results (marks). \label{tab:Summary}}
	
	{\setlength{\tabcolsep}{0.5em}
	\begin{tabular}{lcccccccc}
		\toprule
		\multirow{2}{*}{\emph{Model}} & \multicolumn{4}{c}{Elasticity} & \multicolumn{4}{c}{Quasi-linear viscoelasticity (QLV)} \\
	     & \multicolumn{2}{c}{exponential} & \multicolumn{2}{c}{polynomial} & \multicolumn{2}{c}{exponential} & \multicolumn{2}{c}{polynomial} \\
	    \midrule
		\emph{Loading} & ramp & sine & ramp & sine & ramp & sine & ramp & sine \\
		\midrule
		\emph{Exact theory} & yes $\text{\ding{51}}_\eqref{ShockElastExp}$ & yes $\text{\ding{51}}_\eqref{ShockElastExp}$ & yes $\text{\ding{51}}_\eqref{ChockaShockRamp}$ & yes $\text{\ding{51}}_\eqref{ChockaShock}$ & --- & --- & --- & --- \\[3pt]
		\emph{Acceleration waves} & yes $\text{\ding{51}}_\eqref{BernoulliBlowup}$ & yes $\text{\ding{51}}_\eqref{BernoulliBlowup}$ & no $\text{\ding{55}}_\eqref{BernoulliBlowup}$ & no $\text{\ding{55}}_\eqref{BernoulliBlowup}$ & y/n $\text{\ding{55}}^*_\eqref{BernoulliBlowup}$ & y/n $\text{\ding{55}}^*_\eqref{BernoulliBlowup}$ & no $\text{\ding{55}}^*_\eqref{BernoulliBlowup}$ & no $\text{\ding{55}}^*_\eqref{BernoulliBlowup}$\\[3pt]
		\emph{Slow scale} & yes $\text{\ding{51}}_\eqref{BurgersDistRamp}$ & yes $\text{\ding{51}}_\eqref{BurgersDistRamp}$ & yes $\text{\ding{55}}_\eqref{BurgersDistRamp}$ & yes $\text{\ding{55}}^*_\eqref{BurgersDistSin}$ & y/n $\text{\ding{55}}^*_\eqref{BurgersDistRamp}$ & y/n $\text{\ding{55}}^*_\eqref{BurgersDistRamp}$ & yes $\text{\ding{55}}^*_\eqref{BurgersDistRamp}$ & y/n $\text{\ding{55}}^*_\eqref{BurgersDistSin}$ \\
		\bottomrule 
		\multicolumn{9}{l}{ \footnotesize{$\text{\ding{51}}$ Correct, $\text{\ding{55}}$ Incorrect, $\text{\ding{55}}^*$ Accuracy potentially restricted to a given range of loading signals.} }
	\end{tabular}}
\end{table}

The paper is organised as follows. Section~\ref{sec:Prelim} presents the equations of motion, which are written as a first-order system of balance laws. In Section~\ref{sec:Accel}, the speed and evolution of acceleration waves are analysed theoretically. Section~\ref{sec:Slow} presents the derivation of a damped Burgers-like equation based on a slow scale assumption, as well as the corresponding shock formation conditions. Section~\ref{sec:Shock} benchmarks these shock formation estimates using a dedicated finite volume scheme. Conclusions and prospects are detailed in Section~\ref{sec:Conclu}. In the Appendix~\ref{app:Linear}, we derive the linear solution. Finally, the Appendix~\ref{app:Elastic} covers exact shock formation estimates obtained in the elastic case \cite{chu64,chockalingam20}.

\section{Governing equations}\label{sec:Prelim}

Similarly to Chu \cite{chu64} (see also Chockalingam and Cohen \cite{chockalingam20}), we consider a homogeneous and isotropic semi-infinite medium whose reference configuration is described by the Lagrangian coordinates ${\bm X} = (X, Y, Z)$ with $-\infty < X , Z < +\infty$ and $Y \geqslant 0$. The body is subjected to a shearing motion $\bm{X}\mapsto \bm{x}$ described by the position vector $\bm{x} = \bm{X} + u(Y,t)\bm{e}_1$ in the deformed configuration, where $\lbrace \bm{e}_1, \bm{e}_2, \bm{e}_3 \rbrace$ is an orthonormal basis of the Cartesian coordinate system. This deformation results from a causal and continuous shearing velocity $\bm{v} = V(t) \bm{e}_1$ applied on the surface $Y = 0$, where $\bm{v} = \dot{\bm x} = v(Y,t)\bm{e}_1$ is the particle velocity and the overdot denotes the material time derivative $\partial_t$.
The simple shear deformation yields the deformation gradient tensor \cite{holzapfel00}
\begin{equation}
	\bm{F} = \frac{\partial{\bm{x}}}{\partial{\bm{X}}} = \bm{I} + \text{Grad}\, \bm{u} = \begin{bmatrix}
		1 & \gamma & 0 \\
		0 & 1 & 0 \\
		0 & 0 & 1
	\end{bmatrix} 
	\label{F}
\end{equation}
where $\gamma = \partial_Y u$ is the shear strain. Here, $\bm{u} = \bm{x} - \bm{X}$ is the displacement field, $\bm{I} = [\delta_{ij}]$ is the second-order identity tensor, and $\text{Grad}$ denotes the gradient operator with respect to the material coordinates ${\bm{X}}$. Various strain tensors are defined as functions of $\bm F$, such as the left and right Cauchy--Green tensors $\bm{B} = \bm{F}\bm{F}^\transpose$ and $\bm{C} = \bm{F}^\transpose \bm{F}$.

In \emph{incompressible} materials, the constraint of no volume dilatation $J = \det\bm{F} \equiv 1$ is prescribed at all times. The strain energy function $W$ depends on the invariants $I_1 = \text{tr}\, \bm{B}$ and $I_2 = \tfrac12 \big(I_1^2 - \text{tr} (\bm{B}^2)\big)$ only, which both equal $3+\gamma^2$ in simple shear \eqref{F}. 
Within the Fung--Simo theory of incompressible Quasi-Linear Viscoelasticity (QLV) \cite{depascalis14,berjamin21}, the total second Piola--Kirchhoff stress is expressed as
\begin{equation}
	\bm{S} = -p \bm{C}^{-1} + {\bm S}^\text{e} - \sum_{k=1}^n \bm{S}^\text{v}_k \, , \qquad 
	\label{FungMemStress}
\end{equation}
where $p = p(Y,t)$ is the arbitrary Lagrange multiplier accounting for the incompressibility constraint.
The elastic stress response $\bm{S}^\text{e} = 2\, \partial W/\partial\bm{C}$ is expressed as
\begin{equation}
	\bm{S}^\text{e} = 2 \left(W_1 + I_1 W_2\right) \bm{I} - 2W_2 \bm{C} \, ,
	\label{Elast}
\end{equation}
where $W_i$ is shorthand for the partial derivative $\partial W/\partial I_i$ evaluated at $(I_1, I_2)$.
The tensors
\begin{equation}
	\bm{S}^\text{v}_k = g_k \int_0^t \big(1-\text{e}^{-(t-s)/\tau_k}\big)\, \dot {\bm S}^\text{e}_\text{D}(s)\, \text d s = \frac{g_k}{\tau_k} \int_0^t \text{e}^{-(t-s)/\tau_k}\, {\bm S}^\text{e}_\text{D}(s)\, \text d s 
	\label{FungMemVar}
\end{equation}
are \emph{memory variables}, where the deviatoric elastic stress ${\bm S}^\text{e}_\text{D} = {\bm S}^\text{e} - \tfrac13 ({\bm S}^\text{e} :\bm{C}) \bm{C}^{-1}$ satisfies
\begin{equation}
	{\bm S}^\text{e}_\text{D} = 2 \left(W_1 + I_1 W_2\right) \bm{I} - 2W_2 \bm{C} - \tfrac23 \left(I_1 W_1 + 2 I_2 W_2\right) \bm{C}^{-1}
\end{equation}
and the colon denotes double contraction. The memory variables satisfy the linear evolution equation
\begin{equation}
	\tau_k \dot{\bm S}^\text{v}_k = g_k {\bm S}^\text{e}_\text{D} - \bm{S}^\text{v}_k
	\label{FungEvol}
\end{equation}
with parameters $g_k$, $\tau_k$ for each viscoelastic relaxation mechanism.

In what follows, we restrict the analysis to $n=1$ relaxation mechanism. Thus, we denote the memory variable as $\bm{S}^\text{v} = \bm{S}^\text{v}_1$ and the viscoelastic parameters as $g=g_1$, $\tau=\tau_1$. For a single relaxation mechanism, one notes in passing that the constitutive law \eqref{FungMemStress}-\eqref{FungEvol} can be written in rate form as follows (see related models in \cite{saccomandi21})
\begin{equation}
    \bm{S} = -q \bm{C}^{-1} + {\bm S}_\text{D}, \qquad
    {\bm S}_\text{D} + \tau\dot{\bm S}_\text{D} = \left(1-g\right) {\bm S}^\text{e}_\text{D} +  \tau\dot{\bm S}^\text{e}_\text{D} ,
    \label{StressRate}
\end{equation}
where $q = p - \tfrac13 \bm{S}^\text{e}:\bm{C}$ and ${\bm S}_\text{D} = {\bm S}^\text{e}_\text{D} - \bm{S}^\text{v}$ (this alternative form of the constitutive law is not used here).
The motion is governed by the momentum balance equation $\rho \dot{\bm v} = \text{Div}\, \bm{P}$ where the mass density $\rho$ is a constant, and the first Piola--Kirchhoff stress tensor $\bm{P} = \bm{F}\bm{S}$ is deduced from Eq.~\eqref{FungMemStress}. These expressions are consistent with the componentwise definition $[\text{Div}\, \bm{P}]_i = P_{ij,j}$ of the divergence computed with respect to $\bm X$, and summation over repeated indices is performed \cite{holzapfel00}.

Upon inspection of the components of the equation of motion, we find that the pressure field $p$ equilibrates compressive tractions. The shearing motion is governed by a system of balance laws of the form
\begin{equation}
	\partial_t {\bf q} + \partial_Y {\bf f} ({\bf q}) = {\bf g} ({\bf q}) ,\qquad {\bf q} = \begin{bmatrix}\gamma\\
		v\\
		r\\
		s
	\end{bmatrix} , \qquad
	{\bf f}({\bf q}) = \begin{bmatrix}
		-v \\
		-\sigma/\rho\\
		0 \\
		0
	\end{bmatrix} , \qquad
	{\bf g}({\bf q}) = \frac{1}{\tau} \begin{bmatrix}
		0 \\
		0 \\
		g [S^\text{e}_\text{D}]_{12} - r\\
		g [S^\text{e}_\text{D}]_{22} - s
	\end{bmatrix} ,
	\label{SystHypVect}
\end{equation}
where $r=[S^\text{v}]_{12}$, $s=[S^\text{v}]_{22}$ are components of the memory variable $\bm{S}^\text{v}$, and the stress components are given by
\begin{equation}
    \sigma = P_{12} = \left(2 W_1 + 2 W_2 - s\right) \gamma - r
    \qquad\text{and}\qquad
	\begin{aligned}
		&[S^\text{e}_\text{D}]_{12} = -2 W_2 \gamma + 2 (W_1 + 2 W_2) \big( \gamma + \tfrac13 \gamma^3\big) ,\\
		&[S^\text{e}_\text{D}]_{22} = -\tfrac23 (W_1 + 2 W_2) \gamma^2 .
	\end{aligned}
	\label{SystStress}
\end{equation}
The material is assumed initially at rest, and a compatible velocity field $V(t)$ is imposed at the boundary $Y=0$:
\begin{equation}
	{\bf q}(Y,0) = {\bf 0}, \qquad v(0,t) = V(t) .
	\label{BCond}
\end{equation}
As long as the memory variables $r$, $s$ remain equal to zero (i.e., if $g =0$ or $\tau \to + \infty$), we recover the exact same equations as in Chockalingam and Cohen \cite{chockalingam20}.

For later use, we introduce also the Jacobian matrices
\begin{equation}
	{\bf A}({\bf q}) = \begin{bmatrix}
		0 & -1 & 0 & 0 \\
		-\tfrac1{\rho}\tfrac{\partial \sigma}{\partial\gamma} & 0 & \tfrac1{\rho} & \tfrac{\gamma}{\rho}\\
		0 & 0 & 0 & 0 \\
		0 & 0 & 0 & 0 
	\end{bmatrix} = \frac{\partial {\bf f}}{\partial {\bf q}} , \qquad
	{\bf B}({\bf q}) = \frac{1}{\tau} \begin{bmatrix}
	    0 & 0 & 0 & 0 \\
	    0 & 0 & 0 & 0 \\
		g\, \partial_\gamma [S^\text{e}_\text{D}]_{12} & 0 & -1 & 0 \\
		g\, \partial_\gamma [S^\text{e}_\text{D}]_{22} & 0 & 0 & -1
	\end{bmatrix} = \frac{\partial {\bf g}}{\partial {\bf q}}
	\label{SystJac}
\end{equation}
deduced from Eq.~\eqref{SystHypVect}. The matrix ${\bf A}({\bf q})$ is diagonalisable with real eigenvalues $\lbrace 0, \pm c \rbrace$ where $\rho c^2 = \partial_\gamma \sigma$ defines the speed of sound $c$, provided that $\partial_\gamma\sigma$ is positive. The matrix ${\bf B}({\bf q})$ is diagonalisable with real eigenvalues $\lbrace 0, -1/\tau \rbrace$ too.

Illustrations will be provided for sinusoidal signals of the form
\begin{equation}
	V(t) = \frac{A}{\Omega} \sin(\Omega t) \left(\operatorname{H}(t) - \operatorname{H}\big(t-k\tfrac{2\pi}\Omega\big)\right) ,
	\label{Sign}
\end{equation}
where $A$ is the initial slope (acceleration), $\Omega\geqslant 0$ is the characteristic angular frequency, $k>0$ is the number of periods, and $\operatorname{H}$ denotes the Heaviside step function. The limit $\Omega \to 0$ yields the ramp function $V(t) = A t \operatorname{H}(t)$.

\begin{table}
	\centering
	\caption{Typical parameter values for brain tissue used in the figures and simulations. The linear viscoelastic parameter values are taken from Rashid et al.~\cite{rashid13}, while the parameters of nonlinearity are consistent with Refs.~\cite{comellas20,berjamin21b}. \label{tab:Param}}
	
	\begin{tabular}{ccccccccc}
		\toprule
		$\rho$ ($\text{kg}/\text{m}^3$) & $\mu$ (kPa) & $C_1$ (kPa) & $C_2$ (kPa) & $g$ & $\tau$ (s) & $\alpha$ & $\beta$ & $b$ \\
		$10^3$ & $4.9$ & $2.45$ & $0$ & $0.306$ & $0.011$ & $1.57$ & $2.2$ & $2.57$ \\
		\bottomrule
	\end{tabular}
\end{table}

\begin{figure}
	\begin{minipage}{0.49\textwidth}
		\centering
		(a)
		
		\includegraphics{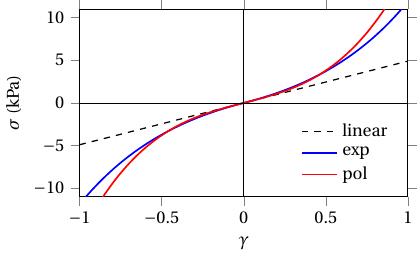}
	\end{minipage}\hfill
	\begin{minipage}{0.49\textwidth}
		\centering
		(b)
		
		\includegraphics{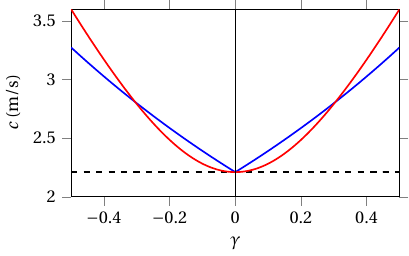}
	\end{minipage}
	
	\caption{Elastic response (limit $g=0$, $\tau\to +\infty$). Evolution of the shear stress (a) and of the shear-wave speed (b) with respect to the shear strain for the exponential and polynomial models \eqref{W} using the parameters of Table~\ref{tab:Param}. The dashed line marks the linear law $\sigma = \mu\gamma$. \label{fig:Param}}
\end{figure}

We will consider materials described by the strain energy functions $W$ expressed by
\begin{equation}
	W^\text{exp} = \frac{\mu}{\alpha^2} \left(\text{e}^{\alpha \sqrt{|I_1 - 3|}} - \alpha \sqrt{|I_1 - 3|} - 1\right) ,\qquad
	W^\text{pol} = C_1 \left(I_1 - 3 + \tfrac12\beta (I_1 - 3)^2\right) + C_2 \left(I_2 - 3\right) .
	\label{W}
\end{equation}
The exponential function $W^\text{exp}$ involves the shear modulus $\mu>0$ and a dimensionless parameter $\alpha > 0$. The polynomial function $W^\text{pol}$ is a combined Yeoh and Mooney--Rivlin model where $\mu = 2(C_1 + C_2)$ is the shear modulus, the coefficients $C_1$, $C_2 > 0$ are the Mooney parameters, and $\beta > 0$ is the Yeoh coefficient (dimensionless). When $\beta$, $C_2$ equal zero, neo-Hookean solids are recovered. Using these expressions, the partial derivatives of the strain energy read
\begin{equation}
	W^\text{exp}_1 = \frac{\mu}{2\alpha |\gamma|} \left(\text{e}^{\alpha |\gamma|} - 1\right) , \quad W^\text{exp}_2 = 0 , \qquad\qquad
	W^\text{pol}_1 = C_1 \left(1 + \beta \gamma^2\right) , \quad W^\text{pol}_2 = C_2 .
	\label{WDer}
\end{equation}
Typical values of the parameters are given in Table~\ref{tab:Param}, and corresponding elastic shear stresses $\sigma$ and wavespeeds $c$ deduced from \eqref{ExpCalc}-\eqref{PolCalc} with $g=0$, $\tau\to +\infty$ are shown in Figure~\ref{fig:Param}. Despite the stress-strain curves looking similar, a noteworthy feature of the exponential model is the singularity of the wavespeed's derivative at zero strain.

\section{Acceleration waves}\label{sec:Accel}

Let us analyse the propagation of \emph{acceleration waves}. Here, the primary field $\bf q$ is continuous across the surface $\xi(\bm{X}, t) = 0$ with $\xi = Y - \eta(t)$ and $\eta$ denoting the position of the acceleration front, but the normal derivative $\partial_\xi {\bf q}$ may be discontinuous. Typically, such solutions represent situations in which the field variables experience a brutal change of slope; for instance, a boundary-value problem of the present form \eqref{BCond} with ramp-type or sinusoidal boundary data \eqref{Sign}.

\subsection{Speed}

We follow the steps in Sec.~8.4 of M{\"u}ller and Ruggeri \cite{ruggeri98}, and assume that the wave propagates into a domain where the primary field ${\bf q}$ is in a stress-free equilibrium state $\bar{\bf q} = {\bf 0}$, so that ${\bf g}(\bar{\bf q})$ vanishes. Similar derivations are also found in the text by Coleman and Gurtin \cite{coleman65}, and in related works \cite{schuler73}.

The jumps $\llbracket \cdot \rrbracket$ of the partial derivatives across the moving surface are related to those of the normal derivative $\partial_Y {\bf q}$ according to $\llbracket \partial_t {\bf q} \rrbracket = -\lambda\, \llbracket \partial_\xi {\bf q} \rrbracket$ and $\llbracket \partial_Y {\bf q} \rrbracket = \llbracket \partial_\xi {\bf q} \rrbracket$, where the wave speed satisfies $\lambda = \partial_t \eta$. Therefore, by computing the jump of Eq.~\eqref{SystHypVect} and using the continuity requirement $\llbracket {\bf q} \rrbracket = {\bf 0}$, we find
\begin{equation}
	\left({\bf A}(\bar{\bf q}) - \lambda\, {\bf I}\right) \llbracket \partial_t {\bf q} \rrbracket = {\bf 0}
	\label{Acceleration}
\end{equation}
along the wavefront. Hence, $\lambda$ is an eigenvalue of ${\bf A}(\bar{\bf q})$, and the jump vector $\llbracket \partial_t {\bf q} \rrbracket$ belongs to the corresponding eigenspace. In other words, $\llbracket \partial_t {\bf q} \rrbracket = \Pi\, {\bf r}(\bar{\bf q})$ is proportional to a basis vector ${\bf r}(\bar{\bf q})$ of this eigenspace.

Given that the present boundary problem involves waves propagating towards increasing $Y$, we restrict the presentation of the eigenspaces of ${\bf A}({\bf q})$ to non-negative eigenvalues. Using Eq.~\eqref{SystJac}, the relevant left- and right eigenvectors ${\bf l}({\bf q})$, ${\bf r}({\bf q})$ corresponding to the only suitable eigenvalue $+c$ are given by
\begin{equation}
	{\bf l}({\bf q}) = \frac12\begin{bmatrix}
		-c\\
		1\\
		1/(\rho c)\\
		\gamma/(\rho c)
	\end{bmatrix} , \qquad
	{\bf r}({\bf q}) = \begin{bmatrix}
		-1/c\\
		1\\
		0\\
		0
	\end{bmatrix} , \qquad
	c = \sqrt{ \frac1\rho\frac{\partial \sigma}{\partial \gamma} } = c({\bf q})
	\label{Speed}
\end{equation}
with the normalisation ${\bf l}^\transpose {\bf r} = 1$. Here, we have scaled ${\bf r}({\bf q})$ in such a way that its velocity entry equals unity (dimensionless). Thus, the eigenvectors' components ${\rm l}_i$, ${\rm r}_i$ have same dimension as $c/{\rm q}_i$, ${\rm q}_i/c$, respectively. The requirement that ${\bf A}({\bf q})$ is diagonalisable with real wave speeds leads to the hyperbolicity condition.

\begin{condition}[Hyperbolicity]
	The quasi-linear first-order system \eqref{SystHypVect} is hyperbolic provided ${\bf q} = [\gamma, v, r, s]^\transpose$ satisfies
	\begin{equation}
		\rho c^2 = \frac{\partial \sigma}{\partial \gamma} = \frac{\partial}{\partial \gamma} \left( 2(W_1+W_2)\gamma \right) - s > 0 ,
		\label{Modulus}
	\end{equation}
	where the coefficients $W_i$ are functions of the shear strain $\gamma$. For the exponential and polynomial constitutive laws \eqref{W}-\eqref{WDer}, hyperbolicity imposes that
	\begin{equation}
		\frac{\partial \sigma^\textup{exp}}{\partial \gamma} = \mu \, \textup{e}^{\alpha |\gamma|} - s > 0  ,
		\qquad
		\frac{\partial \sigma^\textup{pol}}{\partial \gamma} = \mu\left(1 + (b \gamma)^2\right) - s > 0 ,
	\end{equation}
	where $b = \sqrt{6 \beta C_1/\mu} > 0$ is dimensionless.
\end{condition}

\begin{remark}
	The exponential and polynomial constitutive laws \eqref{W}-\eqref{WDer} are not asymptotically equivalent in the limit of small shear strains, see Fig.~\ref{fig:Param}. In fact,
	\begin{equation}
		\frac{\partial \sigma^\textup{exp}}{\partial \gamma} \; \underset{(\gamma\to 0)}{\simeq} \; \mu \left(1 + \alpha |\gamma| +\tfrac12 (\alpha |\gamma|)^2\right) - s
		\label{ExpSeries}
	\end{equation}
	includes a linear term in $|\gamma|$, whereas $\partial_\gamma \sigma^\textup{pol}$ is purely quadratic in $\gamma$. Nevertheless, if we impose that elastic shock times obtained with ramp-type loading match (see Appendix~\ref{app:Elastic}), then the coefficients are linked through $\alpha \approx 0.612\, b$. The parameter $\alpha$ in Table~\ref{tab:Param} has been set accordingly using the value of $\beta$ in Ref.~\cite{berjamin21b}.
\end{remark}

With these observations in mind, one notes that the \emph{wave amplitude} $\Pi = \llbracket \partial_t v \rrbracket$ is an acceleration (in $\text{m}\,\text{s}^{-2}$). From $\bar{\bf q} = {\bf 0}$, Eqs.~\eqref{Speed}-\eqref{Modulus} then tell us that the speed $\lambda = c(\bar{\bf q})$ of acceleration waves equals the speed $c_0 = \sqrt{ \mu/\rho }$ of linear elastic shear waves, where $\mu = 2\, (W_1 + W_2) |_{\gamma = 0}$ is the shear modulus. The expression of ${\bf r}$ shows that the jumps satisfy $\llbracket \partial_t \gamma \rrbracket = -\Pi/c_0$, and that the derivatives $\partial_t r$, $\partial_t s$ of the memory variables are continuous across the wavefront.

\subsection{Evolution}

Upon spatial differentiation of the scalar product between ${\bf l}({\bf q})$ and \text{Eq.}~\eqref{SystHypVect}, computation of the jumps shows that the wave amplitude satisfies the following Bernoulli differential equation \cite{ruggeri98}
\begin{equation}
	\frac{\text d}{\text d t}\Pi = \Omega_2 \Pi^2 - \Omega_1 \Pi \, ,
	\label{Bernoulli}
\end{equation}
where $\frac{\text d}{\text d t} = \partial_t + c_0\,\partial_Y$ denotes the directional derivative along the curve $Y = c_0 t$. The coefficients are obtained by evaluating the following expressions at the constant equilibrium state ${\bf q} = {\bf 0}$:
\begin{equation}
    \Omega_1 = -{\bf l} \cdot {\bf B}\, {\bf r}  = \frac{g}{2\mu\tau} \frac{\partial [S^\text{e}_\text{D}]_{12}}{\partial \gamma} =  \frac{g}{2\tau} ,
    \qquad\qquad
	\Omega_2 = \frac1{c_0} \frac{\partial c}{\partial {\bf q}} \cdot {\bf r} = -\frac{1}{c_0^2} \frac{\partial c}{\partial \gamma} = -\frac{2}{\mu c_0} (W'_1 + W'_2) ,
	\label{BernoulliCoeff}
\end{equation}
where $W'_i = \partial_\gamma W_i$ are functions of $\gamma$, and where we have used Eqs.~\eqref{SystHypVect}-\eqref{SystJac} and \eqref{Speed}. The characteristic frequency of decay $\Omega_1 = g/(2\tau) \geqslant 0$ depends on the viscoelastic parameters only, whereas the nonlinearity parameter $\Omega_2$ depends on the hyperelastic strain energy.

A well-known analytical solution to Eq.~\eqref{Bernoulli} yields the time-evolution of the jump amplitude
\begin{equation}
	\Pi(t) = \frac{A\, \text{e}^{-\Omega_1 t}}{1 - A \frac{\Omega_2}{\Omega_1} \big(1 - \text{e}^{-\Omega_1 t}\big)},
	\label{BernoulliSol}
\end{equation}
as the wave propagates, where we have used the boundary condition $\Pi(0) = A$ from Eq.~\eqref{BCond} with the signal \eqref{Sign}. Thus, the acceleration wave collapses into a shock wave in finite time at
\begin{equation}
	t^* = -\frac{1}{\Omega_1} \ln \left(1 - \frac{\Omega_1}{A\Omega_2}\right) >0, \qquad
	Y^* = c_0 t^* > 0 ,
	\label{CriticalTime}
\end{equation}
provided that $A$ and $\partial_\gamma c |_{{\bf q} = {\bf 0}}$ have opposite signs. In the elastic case (Appendix~\ref{app:Elastic}), the conditions \eqref{ShockElastAcc} of shock formation along the wavefront are recovered by setting $g=0$, $\tau \to +\infty$ in the above expressions (i.e., by setting $\Omega_1\to 0$).

\begin{figure}
    \begin{minipage}{0.49\textwidth}
        \centering
        (a)
        
        \includegraphics{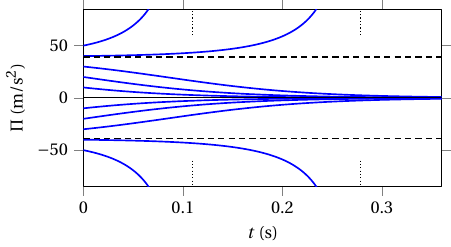}
    \end{minipage}\hfill
    \begin{minipage}{0.49\textwidth}
        \centering
        (b)
        
        \includegraphics{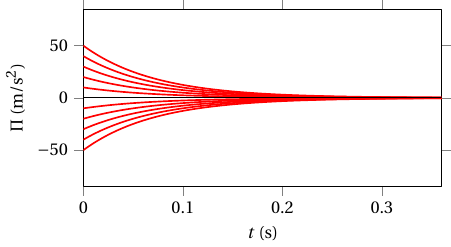}
    \end{minipage}
	\caption{Evolution of the amplitude of nonlinear acceleration shear waves \eqref{BernoulliSol} for initial amplitudes $A$ ranging from $-50$ to $50$ m/s\textsuperscript{2}. (a) Exponential model, where dashed lines mark the critical amplitudes $\pm 2 \Omega_1 c_0 / \alpha$ and dotted lines mark the blow-up times \eqref{BernoulliBlowup}; (b) Polynomial model. \label{fig:Evol}}
\end{figure}

For sake of illustration, let us go back to the exponential and polynomial models described by the strain energy functions $W^\text{exp}$, $W^\text{pol}$ of Eq.~\eqref{W}. With these expressions, the nonlinearity parameter $\Omega_2$ becomes
\begin{equation}
	\Omega_2^\text{exp} = \frac{\alpha}{2c_0} \frac{A}{|A|} ,
	\qquad \Omega_2^\text{pol} = 0 ,
	\label{BernoulliNL}
\end{equation}
where we have used the fact that $-\llbracket \partial_t \gamma \rrbracket$ and $\Pi$, $A$ have the same sign. Therefore, acceleration waves blow up at the critical time
\begin{equation}
	(t^*)^\text{exp} = -\frac{1}{\Omega_1} \ln \left(1 - \Omega_1 \frac{2 c_0}{\alpha|A|}\right) ,
	\qquad (t^*)^\text{pol} \in \emptyset .
	\label{BernoulliBlowup}
\end{equation}
As can be seen from Eq.~\eqref{BernoulliBlowup}, acceleration waves do not collapse into shocks for the polynomial model, where the evolution of $\Pi$ is the same as for linear viscoelastic solids (exponential decay in Fig.~\ref{fig:Evol}b). Nevertheless, as shown later on, shock waves could still be generated beyond the first edge of the impinging wavefront.

For the exponential model, acceleration waves lead to shocks in finite time \eqref{BernoulliBlowup} provided that the wave amplitude satisfies $|A| > 2 \Omega_1 c_0 / \alpha$. Contrary to the elastic case $\Omega_1 \to 0$ for which acceleration jumps always blow up in finite time at the exact shock time \eqref{ShockElastExp}, this phenomenon happens only beyond a critical wave amplitude in the viscoelastic case. This feature is illustrated in Figure~\ref{fig:Evol}a where the evolution of $\Pi$ in time is displayed for several wave amplitudes $A=\Pi(0)$ and the parameters in Table~\ref{tab:Param}. The inequality $-\ln (1-x) > x$ then shows that this viscoelastic shock time $(t^*)^\text{exp}$ is necessarily larger than the elastic shock time $\frac{2c_0}{\alpha|A|}$. The shock distance $Y^*$ is obtained by multiplication of the shock time by the velocity $c_0$, see Eq.~\eqref{CriticalTime}. Note that large amplifications of the wavefront slope occur exclusively near the blow-up time, which was also found in the elastic case for more general problems \cite{chockalingam20}.

Here we find the critical amplitude $2 \Omega_1 c_0 / \alpha \approx 39.2~\text{m/s}^2 \approx 4\, \text{g}_0$, where $\text{g}_0 \approx 9.81~\text{m/s}^2$ is the standard acceleration due to gravity. In comparison, high-level karate athletes routinely reach punching accelerations in the $4$-$5~\text{g}_0$ range \cite{loturco14}. With an initial acceleration of $4.5~\text{g}_0$, the blow-up time of acceleration waves \eqref{BernoulliBlowup} is $(t^*)^\text{exp} \approx 0.15$~s, leading to infinite values of the acceleration field within $(Y^*)^\text{exp} \approx 35$~cm of the boundary. In order to get a better understanding of the transfer of boundary accelerations to the brain in traumatic situations, it is therefore of crucial importance that constitutive behaviours and related parameters are estimated in dynamic configurations \cite{jiang15}. In Section~\ref{sec:Shock}, we show that the value of the critical acceleration $2 \Omega_1 c_0 / \alpha$ should be used with caution, as shocks might form despite decay of acceleration waves.

\section{Slow scale}\label{sec:Slow}

Shock formation can be studied analytically in a similar fashion to Zabolotskaya et al. \cite{zabo04}. In fact, let us introduce the following \emph{scaling} defined by the change of variables $\lbrace \tilde Y = \epsilon^2 Y, \tilde t = t-Y/c_0, u= \epsilon \tilde u, r = \epsilon^3 \tilde r, s = \epsilon^3 \tilde s \rbrace$ where $\epsilon$ is a small positive parameter, and $u = u (Y, t)$, $\tilde u = \tilde u (\tilde Y, \tilde t)$, etc. Furthermore, we assume that $1/\tau$ is of order $\epsilon^2$, and that $\alpha$ is of order $\epsilon$. This Ansatz is then substituted in the equations of motion \eqref{SystHypVect}. At leading (cubic) order in $\epsilon$, the motion of exponential and polynomial QLV solids is governed by a scalar equation of the form
\begin{equation}
    \epsilon^3\frac{2\mu}{c_0} \partial_{\tilde t\tilde Y} \tilde u = \epsilon^{m+1} \frac{2\rho a}{m+1} \partial_{\tilde t} \left( |\partial_{\tilde t}\tilde u|^m \partial_{\tilde t}\tilde u\right) + \frac{\epsilon^3}{c_0} \partial_{\tilde t} \tilde r
    \qquad\text{with}\qquad
    \frac{\epsilon^3}{c_0} \partial_{\tilde t} \tilde r = -\epsilon\rho \frac{g}{\tau} \partial_{\tilde t} \tilde u , \quad \epsilon^3 \partial_{\tilde t} \tilde s = 0,
    \label{SSinter}
\end{equation}
where the coefficients $a$, $m$ are specified subsequently. Transforming \eqref{SSinter} back to the space coordinate $Y$ and displacement field $u(Y, \tilde t) = \epsilon \tilde u(\tilde Y, \tilde t)$ yields the scalar transport equation
\begin{equation}
    c_0\, \partial_Y v = a \left|v\right|^m\, \partial_{\tilde t} v - \Omega_1 v
    \qquad
	\text{with}\qquad 
	(a, m)^\text{exp} = \left(\frac{\alpha}{2 c_0}, 1\right),\quad
	(a, m)^\text{pol} = \left(\frac{b^2}{2 c_0^2}, 2\right) ,
	\label{Burgers}
\end{equation}
and $\Omega_1 = g/(2\tau)$. Here, we have introduced the velocity $v = \partial_{\tilde t} u$ which is therefore viewed as a function of $(Y,\tilde t)$. Compared to the initial definition $v = \partial_t u$ of the velocity $v(Y,t)$, this is slightly abusive but standard notation. Clearly, a major difference between the exponential and polynomial models is the acoustic nonlinearity. Exponents $m=1$ or $m=2$ correspond to asymptotically quadratic or cubic stress-strain relationships, see remark for Eq.~\eqref{ExpSeries}.

Similarly to Zabolotskaya et al. \cite{zabo04}, the solution to the boundary-value problem at hand follows from the method of characteristics. It can be written in implicit form as
\begin{equation}
    v = V(t_0)\, \text{e}^{-\Omega_1 \theta} ,
    \qquad
    t_0 = \tilde t + \frac{a \left|v\right|^m}{m \Omega_1} \left(\text{e}^{m \Omega_1 \theta} - 1\right)
    \label{BurgersSol}
\end{equation}
with $\tilde t = t - \theta$ and $\theta={Y}/{c_0}$, where the boundary condition \eqref{BCond} was used. Differentiation of \eqref{BurgersSol} with respect to $\tilde t$ then yields the acceleration
\begin{equation}
    \partial_{\tilde t} v = \frac{V'(t_0)\, \text{e}^{-\Omega_1 \theta}}{1 - \frac{\varphi(t_0)}{m\Omega_1}\, \big(1 - \text{e}^{-m\Omega_1 \theta}\big) } ,
    \qquad \varphi(t_0) = a\, \frac{\text d}{\text d t_0} \left|V(t_0)\right|^m
    \label{BurgersAcc}
\end{equation}
where $V'(t_0) = A \cos(\Omega t_0)$ and $\varphi(t_0) = a m V'(t_0) V(t_0) |V(t_0)|^{m-2}$.
Interestingly, the acceleration at the wavefront \eqref{BernoulliSol} is recovered by setting $t_0\to 0$ in Eqs.~\eqref{BurgersSol}-\eqref{BurgersAcc}. For the particular case of ramp signal $\Omega\to 0$, the above solution can be written explicitly using
\begin{equation}
    (t_0)^\text{exp} = \frac{\tilde t}{1 - \frac{a^\text{exp}}{\Omega_1} |A|\, \big(1 - \text{e}^{-\Omega_1 \theta}\big)}\, ,
    \qquad\qquad
    (t_0)^\text{pol} = \tilde t\, \frac{1 - \sqrt{1 - 2\frac{a^\text{pol}}{\Omega_1} A^2 \tilde t\, \big(1 - \text{e}^{-2\Omega_1 \theta}\big)}}{\frac{a^\text{pol}}{\Omega_1} A^2 \tilde t\, \big(1 - \text{e}^{-2\Omega_1 \theta}\big)}\, .
\end{equation}

Similarly to Lax \cite{lax73}, we introduce the directional derivative $\frac{\text d}{\text d\theta} = c_0 \partial_Y - a |v|^m \partial_{\tilde t}$ to rewrite the transport equation \eqref{Burgers} as $\frac{\text d}{\text d\theta}v = -\Omega_1 v$, whose solution is given in Eq.~\eqref{BurgersSol}. Differentiation of \eqref{Burgers} with respect to $\tilde t$ then yields the Bernoulli-type differential equation $\frac{\text d}{\text d\theta}q = a \big(\frac{\text d}{\text d v} |v|^m \big) q^2 - \Omega_1 q$ for the accelerations $q = \partial_{\tilde t} v$ whose solution is shown in Eq.~\eqref{BurgersAcc}. Using this differential equation, one notes that the absolute acceleration $|q|$ is initially growing with $\theta$ provided that the quantity $\varphi = \varphi(t_0)$ defined in Eq.~\eqref{BurgersAcc} satisfies $\varphi > \Omega_1$; otherwise $|q|$ is initially decaying. The absolute acceleration experiences a monotonous evolution with $\theta$ for the exponential model $m=1$ (either monotonously increasing or decreasing). For the polynomial model $m=2$, the quantity $|q|$ experiences a monotonous evolution provided that $\varphi < \Omega_1$ or $\varphi>2\Omega_1$. Otherwise, the absolute acceleration grows to reach a local maximum amplification of magnitude $\Omega_1/\sqrt{\varphi (2\Omega_1-\varphi)}$ at $-2\Omega_1 \theta = \ln(2\Omega_1/\varphi - 1)$.

\begin{figure}
    \begin{minipage}{0.32\textwidth}
        \centering
        (a)
        
        \includegraphics{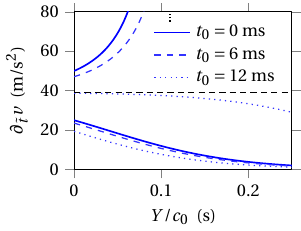}
    \end{minipage}\hfill
    \begin{minipage}{0.33\textwidth}
        \centering
        (b)
        
        \includegraphics{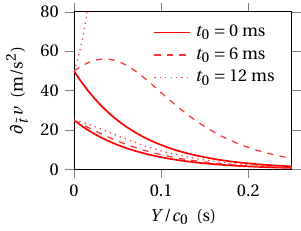}
    \end{minipage}
    \begin{minipage}{0.33\textwidth}
        \centering
        (c)
        
        \includegraphics{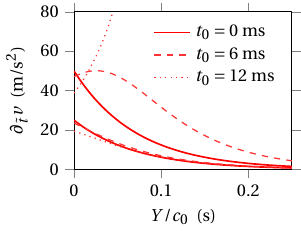}
    \end{minipage}
    \caption{Evolution of the acceleration \eqref{BurgersAcc} along characteristic curves for increasing values of the starting time $t_0$, and fixed amplitudes $A$ equal to 25 or $50~\text{m/s}^2$. (a) Exponential model with sinusoidal loading $\Omega = 57$~rad/s; Polynomial model with (b) ramp loading $\Omega\to 0$ and (c) sinusoidal loading of same frequency as in Fig.~\ref{fig:EvolSS}a.}
    \label{fig:EvolSS}
\end{figure}

The evolution of $q$ with respect to $\theta$ is shown in Fig.~\ref{fig:EvolSS}. For ramp-type signal $\Omega\to 0$, the exponential model produces the same evolution of the acceleration as in Fig.~\ref{fig:Evol}a for all $t_0$. This is no longer true for sinusoidal signal, where increasing values of $t_0$ delay the formation of shocks at fixed amplitude (Fig.~\ref{fig:EvolSS}a). For the polynomial model, increasing values of $t_0$ may introduce a local maximum acceleration or even a vertical asymptote (Figs.~\ref{fig:EvolSS}b-\ref{fig:EvolSS}c), while the case $t_0 \to 0$ of Fig.~\ref{fig:Evol}b yields decaying accelerations.

Accelerations \eqref{BurgersAcc} become unbounded at the shock distance and shock time
\begin{equation}
    Y^* = \inf_{t_0>0} Y^\dagger(t_0), \qquad
    t^* = \inf_{t_0>0} t^\dagger(t_0),
    \label{BurgersDistGen}
\end{equation}
\begin{equation*}
    \text{where}\qquad
    Y^\dagger(t_0) = -\frac{c_0}{m\Omega_1} \ln\left(1 - \frac{m\Omega_1}{\varphi(t_0)}\right) > 0 , \qquad
    t^\dagger(t_0) = \frac{Y^\dagger(t_0)}{c_0} + t_0 - \frac{V(t_0)}{m V'(t_0)} > 0 .
\end{equation*}
Differentiation of $t^\dagger$ with respect to $t_0$ shows that the blow-up distance $Y^\dagger$ and time $t^\dagger$ are not necessarily optimal for the same $t_0$, and this even in the elastic limit $\Omega_1\to 0$ where $Y^\dagger(t_0) = c_0/\varphi(t_0)$. Note however that the equivalence property between minimal shock distances and times was found true in the elastic limit using an exact theory \cite{chockalingam20}.

For ramp-type forcing $\Omega\to 0$ where $\varphi(t_0) = a m |A|^{m} t_0^{m-1}$, we find the minimal shock distances and times
\begin{equation}
    \begin{aligned}
        (Y^*)^\text{exp} &= Y^\dagger(0) , \\
        (t^*)^\text{exp} &= t^\dagger(0) ,
    \end{aligned}
    \qquad\qquad
    (Y^*)^\text{pol} = Y^\dagger(+\infty) , \quad
    (t^*)^\text{pol} = t^\dagger\left(\frac{\Omega_1c_0^2}{b^2A^2} + \sqrt{\left(\frac{\Omega_1c_0^2}{b^2A^2}\right)^2 + \frac{2c_0^2}{b^2A^2}}\right) .
    \label{BurgersDistRamp}
\end{equation}
Note in passing that $(Y^*)^\text{exp}$ and $(t^*)^\text{exp}$ have the same expression as deduced from the acceleration wave analysis \eqref{CriticalTime}-\eqref{BernoulliBlowup}, which is exact \eqref{ShockElastExp} in the elastic limit $\Omega_1\to 0$. In contrast, the slow scale theory \eqref{BurgersDistRamp} produces $(Y^*)^\text{pol}=0$ and $(t^*)^\text{pol}>0$ whereas the acceleration wave analysis \eqref{BernoulliBlowup} predicts no shock formation.

Sinusoidal forcing $\Omega > 0$ yields the same values of $(Y^*)^\text{exp}$, $(t^*)^\text{exp}$ as for ramp signal \eqref{BurgersDistRamp}. If a polynomial solid is subjected to sinusoidal loading, we find
\begin{equation}
    (Y^*)^\text{pol} = Y^\dagger\left(\frac{\pi}{4\Omega}\right) = -\frac{c_0}{2\Omega_1} \ln\left(1 - 2\Omega_1 \frac{2c_0^2 \Omega}{b^2 A^2}\right) .
    \label{BurgersDistSin}
\end{equation}
Except in the elastic limit (Appendix~\ref{app:Elastic}), the computation of $(t^*)^\text{pol}$ requires numerical minimisation of $t^\dagger$. In Eq.~\eqref{BurgersDistSin}, the elastic limit $\Omega_1\to 0$ recovers the shock distance $2 c_0^3 \Omega/(b A)^2$ of Zabolotskaya et al. \cite{zabo04}, Eq.~(27) therein.

It is well-known that Burgers-like equations with relaxation \eqref{Burgers} can exhibit conditional formation of shocks, as found in Eqs.~\eqref{BurgersDistRamp}-\eqref{BurgersDistSin}. Contrary to results deduced from the study of acceleration waves (see discussion in Section~\ref{sec:Accel}), shocks may also form within finite propagation distances for the polynomial model. In the case of ramp signal, shock formation is always possible. In the case of sinusoidal loading \eqref{BurgersDistSin}, shocks can form provided that the loading amplitude satisfies $|A| > 2 \sqrt{\Omega_1 \Omega}\, c_0/b$, and accelerations decay monotonously provided that the loading amplitude satisfies $|A| < \sqrt{2\Omega_1 \Omega}\, c_0/b$. For both constitutive laws, we note that shock distances tend to increase with increasing dissipation $\Omega_1$, whereas they tend to decrease with increasing amplitudes $|A|$ and nonlinearity $\alpha$, $b$.

\begin{remark}
The present scaling procedure is not unique. For instance, assuming both $1/\tau$ and $g$ of order $\epsilon$ yields the same reduced wave equation \eqref{Burgers}. In addition, other asymptotic procedures might lead to model equations of different nature. For further insights, interested readers are referred to Refs.~\cite{fusco84,pucci19}.
\end{remark}

\section{Numerical experiments}\label{sec:Shock}

\subsection{Numerical method}

In order to further investigate shock formation, we go back to the equations of motion \eqref{SystHypVect} and implement a finite volume scheme. The algorithm is briefly summarised hereinafter. In the examples presented later on, the physical domain is assumed unbounded. We consider a finite numerical domain for $Y$ in $[0 , 1]$~m. It is discretised using a regular grid in space with mesh size $\Delta x$ in the $Y$-direction. The coordinates of the nodes are $y_i = i\, \Delta x$, where $0\leqslant i\leqslant N_x$. The total number of nodes is $N_x+1$, where $N_x = 1/\Delta x$ denotes the number of cells. A variable time step $\Delta t = t_{n+1} - t_n$ is introduced. Therefore, $\mathbf{q}(y_i, t_n)$ denotes the solution to \eqref{SystHypVect} at the grid node $i$ and at the $n$th time step. Numerical approximations of the solution are denoted by $\mathbf{q}_i^n \simeq \mathbf{q}(y_i, t_n)$.

The system of balance laws
\eqref{SystHypVect} is integrated explicitly according to Strang's splitting scheme \cite{leveque02, berjamin18}
\begin{equation}
    \mathbf{q}_{i}^{n+1} = \mathcal{H}_b^{\Delta t/2} \mathcal{H}_a^{\Delta t} \mathcal{H}_b^{\Delta t/2}\, \mathbf{q}_{i}^n ,
	\label{SchemaExplCons}
\end{equation}
where the discrete operator $\mathcal{H}_a^{\Delta t}$ corresponds to the integration of the propagation part over one time step, and the discrete operator $\mathcal{H}_b^{\Delta t/2}$ corresponds to the integration of the viscoelastic relaxation part over a half step. These operators are specified as follows:

\begin{itemize}
    \item For the \emph{propagation part}, we integrate numerically the system of conservation laws $\partial_t {\bf q} + \partial_Y {\bf f} ({\bf q}) = {\bf 0}$ over one time step by using a conservative finite volume scheme
    \begin{equation}
        \mathcal{H}_a^{\Delta t}:\qquad
        \mathbf{q}_{i}^{n+1} = \mathbf{q}_{i}^{n} - \frac{\Delta t}{\Delta x} \left({\bf f}_{i+\frac12} - {\bf f}_{i-\frac12}\right) .
        \label{FV}
    \end{equation}
    Its numerical flux
    \begin{equation}
        {\bf f}_{i+\frac12} = \frac12 \left({\bf f}({\bf q}_{i+\frac12}^-) + {\bf f}({\bf q}_{i+\frac12}^+)\right) - \frac12 \left(\int_0^1 |{\bf A}|\big(s{\bf q}_{i+\frac12}^+ + (1-s){\bf q}_{i+\frac12}^-\big)\, \text{d}s\right) \big({\bf q}_{i+\frac12}^+ - {\bf q}_{i+\frac12}^-\big)
        \label{Osher}
    \end{equation}
    is deduced from Osher's Riemann solver, where the matrix-valued function
    \begin{equation}
        |{\bf A}|({\bf q}) = \begin{bmatrix}
    		c & 0 & -1/(\rho c) & -\gamma/(\rho c) \\
    		0 & c & 0 & 0\\
    		0 & 0 & 0 & 0 \\
    		0 & 0 & 0 & 0 
    	\end{bmatrix} , \qquad
    	c = \sqrt{ \frac1\rho\frac{\partial \sigma}{\partial \gamma} } = c({\bf q})
    \end{equation}
    is integrated numerically using the three-point Gauss--Legendre quadrature method \cite{dumbser11}. The vectors ${\bf q}_{i+1/2}^\pm$ are reconstructed field values at the cell interface $y_{i+1/2}$. Here, they are deduced from the `extrapolate-evolve' MUSCL procedure based on MC-limited slopes \cite{toro09}.\footnote{Monotonic Upstream–centred Scheme for Conservation Laws (MUSCL),  Monotonized Central-difference (MC).} The first-order Osher scheme is recovered in the abscence of MUSCL reconstruction, that is if we set ${\bf q}_{i+1/2}^- = {\bf q}_i^n$ and ${\bf q}_{i+1/2}^+ = {\bf q}_{i+1}^n$ in Eq.~\eqref{Osher}.
    One notes that the propagation step \eqref{FV} modifies the variables $\gamma$, $v$ only; memory variables $r$, $s$ are left invariant. It is stable under the Courant--Friedrichs--Lewy (CFL) condition $\text{Co} \leqslant 1$, where $\text{Co} = \max_{i} c({\bf q}_i^n) \frac{\Delta t}{\Delta x}$ is the maximum Courant number over the domain at time $t_n$.
    \item For the \emph{relaxation part}, we integrate analytically the system of differential equations $\partial_t {\bf q}  = {\bf g} ({\bf q})$, which modifies the memory variables $r$, $s$ according to the differential equation \eqref{FungEvol} with $k=1$. The kinematic variables $\gamma$, $v$ are left invariant, i.e. $\bm{S}^\text{e}_\text{D}$ is a constant. Integration of this differential equation over one time step yields the updating formula
    \begin{equation}
        \mathcal{H}_b^{\Delta t}:\qquad
        (\bm{S}^\text{v})_i^{n+1} = \text{e}^{-\Delta t/\tau} \left((\bm{S}^\text{v})_i^{n} + \Delta t\, \frac{g}{\tau} ({\bm S}^\text{e}_\text{D})_i^{n}\right) 
        \label{Relaxation}
    \end{equation}
    from which the memory variables $r = [S^\text{v}]_{12}$, $s = [S^\text{v}]_{22}$ at time $t_{n+1}$ are obtained. The relaxation step \eqref{Relaxation} is unconditionally stable.
\end{itemize}

\noindent
Due to the stability properties of each sub-step, the split scheme \eqref{SchemaExplCons} is stable under the CFL condition $\text{Co} \leqslant 1$. In practice, we set $\text{Co} = 0.95$ in the numerical examples. At each iteration, the value of $\Delta t$ then follows from the definition of the Courant number.

The time-stepping formula \eqref{SchemaExplCons} is used for the interior cells $0\leqslant i\leqslant N_x$. Boundary conditions are taken into account by addition of ghost cells beyond the domain boundaries, where specific operations are performed during the propagation step. Here, an outflow condition is implemented at the boundary $i=N_x$ of the numerical domain. At the boundary $i=0$, a linearised `incoming wave' condition is implemented in elastic cases ($g=0$, $\tau \to +\infty$), while an `oscillating wall' condition is implemented in viscoelastic cases (see Ref.~\cite{berjamin18} and Chapter~7 of LeVeque \cite{leveque02}). These choices have been made based on error measurements with respect to the linear solution of Appendix~\ref{app:Linear}.

\subsection{Validation}

\paragraph{Linear solution.}

For this test, linear visco-elastic behaviour is assumed ($\alpha = \beta = 0$). We consider sinusoidal signals $\Omega > 0$ restricted to $k=1$ period. In order to perform error measurements, we consider a smoother $C^6$ wavelet \cite{berjamin18phd} instead of a pure truncated sinusoid \eqref{Sign}. The fundamental frequency $\Omega$ of the source is set to the frequency $\sqrt{1-g}/\tau$ of maximum dissipation (see Appendix~\ref{app:Linear}). With the values in Table~\ref{tab:Param}, we find that the maximum dissipation factor equals 0.18 at the frequency $12.1$~Hz. The amplitude of the boundary velocity is $A/\Omega \approx 1$~m/s, where $A = 75$~m/s\textsuperscript{2}.

Figure~\ref{fig:LinearTest} compares the numerical solution to the reference solution of Appendix~\ref{app:Linear} in the elastic and viscoelastic cases. The reference solution is obtained analytically in the elastic case, and by Fourier synthesis in the viscoelastic case. These computations were performed using $N_\text{h} = 60\, 000$ Fourier harmonics in total (including negative and positive frequencies), with a frequency step of $10^4/(N_\text{h}\, t) \approx 0.56$~Hz. The implementation of this computation was benchmarked in the elastic case against analytical expressions.

\begin{figure}
    \begin{minipage}{0.49\linewidth}
        \centering
        
        (a)
        
        \includegraphics{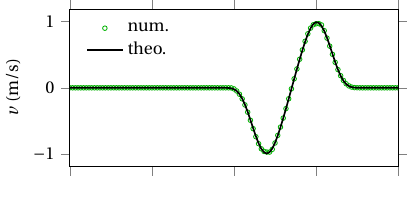}
        \vspace{-1em}
        
        \hspace{-1em}\includegraphics{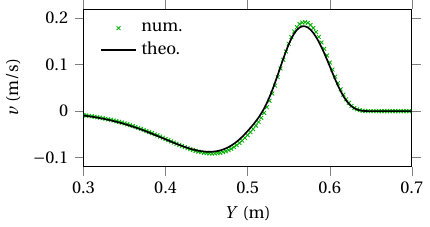}
    \end{minipage}\hfill
    \begin{minipage}{0.49\linewidth}
        \centering
        (b)
        
        \includegraphics{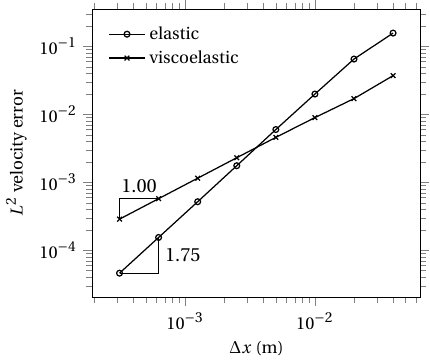}
    \end{minipage}
    
    \caption{Linear solution for a smooth sinusoidal loading. (a) Numerical and theoretical solutions obtained in the elastic case $g=0$, $\tau\to +\infty$ (top) and viscoelastic case (bottom) with parameters from Table~\ref{tab:Param}. (b) Error measurements.}
    \label{fig:LinearTest}
\end{figure}

Waveforms at the time $t \approx 0.3$~s with $N_x = 300$ points are shown in Fig.~\ref{fig:LinearTest}a. In comparison to the elastic case (top), the viscoelastic case (bottom) highlights decaying wave amplitudes, as well as the formation of a low-frequency `tail'. These features illustrate the dissipative and dispersive behaviour of the material. Computational signals have been obtained with 55 grid points per wavelength. Amplitude and phase errors are visible, but these numerical artifacts vanish as the mesh is refined.

Convergence is illustrated in Fig.~\ref{fig:LinearTest}b, where the error in $L^2$-norm increases with the mesh size $\Delta x$. Here, the discretisation includes at least 5 points per wavelength, and at most 585 points per wavelength. The elastic error curve has slope 1.75 in logarithmic coordinates, showing that the scheme is nearly second-order accurate in space and time. However, the experimental order of accuracy is reduced to unity in the viscoelastic case, due to the straightforward implementation of the moving-boundary condition that does not compensate for first-order errors{\,---\,}see Sec.~17.9 of LeVeque \cite{leveque02}. Nevertheless, the scheme is still nearly second-order accurate away from the boundaries (not shown).

\paragraph{Elastic limit, ramp.}

Now we consider nonlinear material behaviour with the parameter values of Table~\ref{tab:Param} and $\Omega_1 \to 0$. For this test, we consider ramp-type signals $\Omega \to 0$ of amplitude $A = \pm 20~\text{m/s}^2$. The numerical solution is discretised using $N_x = 300$ points and run upto a final time of $t\approx 0.14$~s. In the elastic limit, exact shock times can be derived (Appendix~\ref{app:Elastic}), and we find $(t^*)^\text{exp} = (t^*)^\text{pol} \approx 0.141$~s.

\begin{figure}
    \centering
    
    \includegraphics{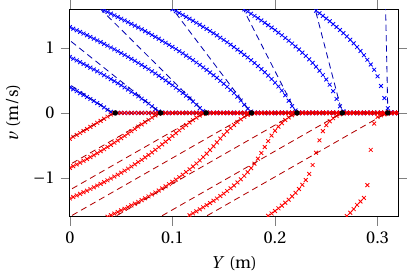}
    
    \caption{Elastic limit. Snapshots of the acceleration waves (dashed lines) and of the numerical results for ramp signal at times ranging from $0.02$~s to $0.14$~s by steps of $0.02$~s as the wave propagates to the right. The exponential model (blue, top) is compared to the polynomial model (red, bottom). \label{fig:RampShockElast}}
\end{figure}

When the exponential model is used, shock times deduced from the acceleration wave and slow scale analyses \eqref{BurgersDistRamp} are exact. This value of $(t^*)^\text{exp}$ is coherent with the simulation results shown in Fig.~\ref{fig:RampShockElast} (top). For the polynomial model (Fig.~\ref{fig:RampShockElast} bottom), the acceleration wave analysis still provides the slope of the impinging wavefront about $v=0$ until it is overtaken by a shock. However, acceleration wave theory cannot inform on the magnitude of slopes beyond the wavefront. The slow scale analysis \eqref{BurgersAcc} with $\Omega_1\to 0$ predicts that shocks form along the loading boundary at infinite times. Nevertheless, infinite accelerations might arise earlier and further away from the boundary \eqref{BurgersDistRamp}. Here shocks should already form at $(t^*)^\text{pol} \approx 0.061$~s and $Y \approx 0.067$~m, but this scenario contradicts both the exact theory and the simulations. Fig.~\ref{fig:ChockaShock} of the Appendix~\ref{app:Elastic} shows that the slow scale estimate is valid for low loading Mach numbers $\text{Ma} = \frac{b A}{c_0 \Omega}$ only, hence it is inaccurate for ramp signals $\Omega\to 0$.

\paragraph{Elastic limit, sine.}

In this test, we consider a truncated sinusoidal signal \eqref{Sign} of $k=1$ period, whose frequency satisfies $\Omega/(2\pi) = 40$~Hz. Using parameter values from Table~\ref{tab:Param} and $\Omega_1\to 0$, we set $N_x = 3\, 200$ in the simulations, so that the numerical solution includes 177 points per wavelength. For the exponential model, we set the initial acceleration to $A = 20$~m/s\textsuperscript{2}, while we use $A = 50$~m/s\textsuperscript{2} for the polynomial model.

\begin{figure}
    \begin{minipage}{0.49\textwidth}
        \centering
        (a)
        
        \includegraphics{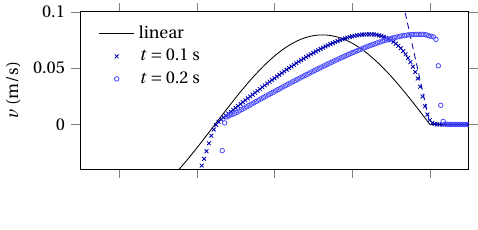}
        \vspace{-2.6em}
        
        \hspace{-0.4em}\includegraphics{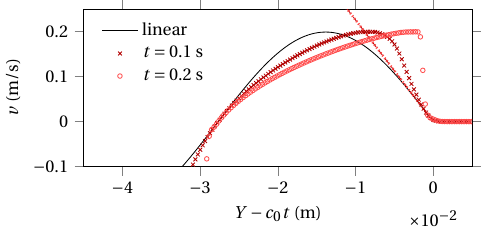}
        
    \end{minipage}\hfill
    \begin{minipage}{0.5\textwidth}
        \centering
        (b)
        
        \includegraphics{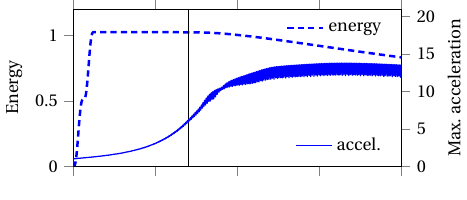}
        \vspace{-1em}
        
        \hspace{-0.16em}\includegraphics{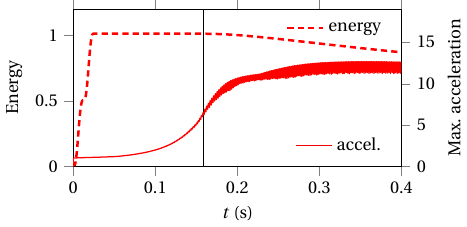}
            
    \end{minipage}
    
    \caption{Elastic limit. (a) Waveforms obtained numerically with the exponential model (top) and the polynomial model (bottom). Broken oblique lines mark the acceleration waves, i.e. the slope of the profile at the wavefront. (b) Normalised energy densities and maximum accelerations. Vertical lines mark the theoretical shock times.}
    \label{fig:ElasticTest}
\end{figure}

Figure~\ref{fig:ElasticTest}a shows the numerical results obtained at increasing times. With both models, waveform steepening is observed, eventually leading to the formation of shock waves. Besides the apparent smoothness of the waveforms, there are some major differences between the exponential model (top) and the polynomial model (bottom). As shown in Section~\ref{sec:Accel}, acceleration waves blow up for the exponential model, providing exact shock time estimates. In contrast, the slope of the wavefront remains invariant for the polynomial model, even though shock waves still form beyond the impinging wavefront. This observation is consistent with the results obtained by Jordan and Puri \cite{jordan05}.

Fig.~\ref{fig:ElasticTest}b represents the time evolution of the energy density $\int_0^1 W(\gamma) + \frac12 \rho v^2 \,\text{d}Y$ for each model, as well as the evolution of the maximum acceleration $\max |\partial_Y \sigma/\rho|$. Both quantities were evaluated numerically, and normalised with respect to their values in the linear case{\,---\,}$\pi\rho c A^2/\Omega^3$ and $A$ respectively. In these diagrams, the faster decay of energy and the large acceleration amplification are a signature of shock formation. In theory, accelerations eventually become infinite; numerically, they take large but finite values locally.

The curves in Fig.~\ref{fig:ElasticTest}b are consistent with the exact shock time estimates $(t^*)^\text{exp} \approx 0.14$~s and $(t^*)^\text{pol} \approx 0.16$~s deduced from Appendix~\ref{app:Elastic} (vertical lines). In this regard, we already mentioned that the slow scale estimate is exact for the exponential model in the elastic limit. For polynomial elastic solids, the slow scale analysis yields $(t^*)^\text{pol} \approx 0.15$~s, see Appendix~\ref{app:Elastic}, which is consistent with the above values. In fact, the Mach number $\text{Ma} \approx 0.23$ has a moderate magnitude, showing that the slow scale estimate is rather accurate for this test (see Fig.~\ref{fig:ChockaShock} of the Appendix~\ref{app:Elastic}).

\subsection{Illustrations}

Let us explore numerically the formation of viscoelastic shocks for various loading programmes \eqref{Sign}. We consider exponential and polynomial QLV material models with the parameters of Table~\ref{tab:Param}. Doing so, we will assess the accuracy of shock time estimates based on the acceleration wave analysis of Sec.~\ref{sec:Accel} and on the scaling of Sec.~\ref{sec:Slow}.

\begin{figure}
    \begin{minipage}{0.49\textwidth}
        \centering
        (a)
        
        \includegraphics{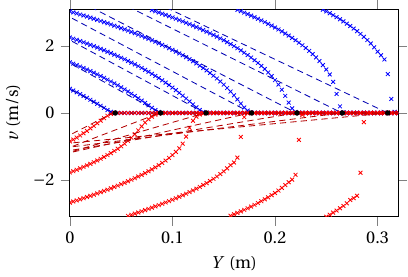}
    \end{minipage}\hfill
    \begin{minipage}{0.5\textwidth}
        \centering
        (b)
        
        \includegraphics{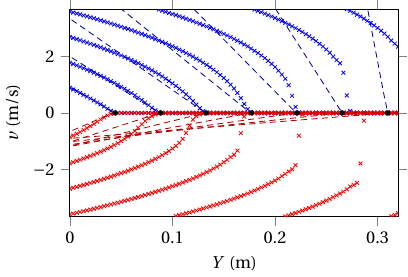}
    \end{minipage}
    
    \caption{Viscoelastic case. Similar figure to Fig.~\ref{fig:RampShockElast} with amplitudes (a) $A = \pm 38~\text{m/s}^2$ and (b) $A = \pm 45~\text{m/s}^2$. \label{fig:RampShock}}
\end{figure}

\paragraph{Ramp signal.} Similarly to the elastic case (Fig.~\ref{fig:RampShockElast}), the numerical solution for ramp signal $\Omega \to 0$ is discretised using $N_x=300$ points. Snapshots of the solution are displayed in Fig.~\ref{fig:RampShock} where time is increased by steps of 0.02~s. For the exponential model, acceleration waves decay in Fig.~\ref{fig:RampShock}a, whereas they grow but remain bounded in Fig.~\ref{fig:RampShock}b. In the first case, the slow scale analysis \eqref{BurgersDistRamp} predicts that shocks never form, whereas in the second case shocks are supposed to form at $(t^*)^\text{exp} \approx 0.147$~s. In both cases, the numerical results show that a shock wave forms within the simulation time, and its speed about $v=0$ is larger than the speed $c_0$ of a smooth solution. In other words, the acceleration wave analysis \eqref{BernoulliBlowup} and the slow scale analysis \eqref{BurgersDistRamp} fail to accurately predict the formation of shocks. Moreover, no critical acceleration behaviour around $2\Omega_1 c_0/\alpha \approx 39.2~\text{m/s}^2$ is seen in the numerical results.

For the polynomial model, the acceleration wave analysis does not inform about shock formation. Besides the minimal shock distance $(Y^*)^\text{pol} = 0$, the slow scale theory predicts that shocks can form earlier on and further away from the boundary. Here, this is supposed to happen at $(t^*)^\text{pol} \approx 0.036$~s, $Y \approx 0.035$~m in Fig.~\ref{fig:RampShock}a and $(t^*)^\text{pol} \approx 0.03$~s, $Y \approx 0.03$~m in Fig.~\ref{fig:RampShock}b, but these estimates are both contradicted by the numerical experiment. In summary, the major difference from the elastic case of Fig.~\ref{fig:RampShockElast} is the loss of accuracy of shock formation predictions for the exponential model.

\paragraph{Sinusoidal signal.} 

\begin{figure}
    \centering
    \includegraphics{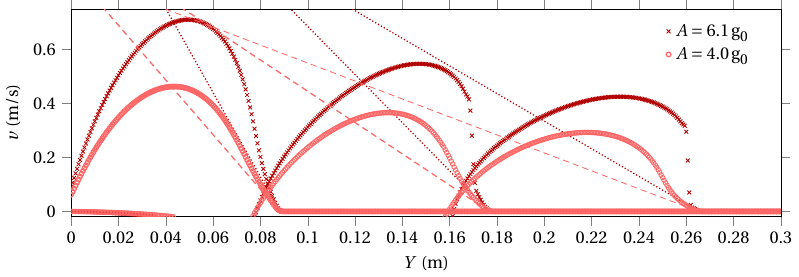}
    \caption{Viscoelastic case. Snapshots of the acceleration waves and of the numerical results for the polynomial model and sinusoidal signal $\Omega/(2\pi) = 12.1$~Hz at times ranging from $0.04$~s to $0.12$~s by steps of $0.04$~s.}
    \label{fig:poly_visc_sin}
\end{figure}

For the case of sinusoidal signals, the elastic limit was already illustrated in Fig.~\ref{fig:ElasticTest}. In the viscoelastic case, let us consider the polynomial model with the parameters of Table~\ref{tab:Param}, amplitudes $A = 39~\text{m/s}^2$ and $A = 60~\text{m/s}^2$. The frequency $\Omega/(2\pi)$ is set at 12.1~Hz (maximum dissipation). The numerical solution is discretised using $N_x = 1\, 600$ (i.e., 293 points per wavelength), and the final time is increased by steps of $0.04$~s until $t\approx 0.12$~s. Numerical results in Fig.~\ref{fig:poly_visc_sin} indicate shock formation within the simulation time for the largest amplitude, whereas the acceleration wave analysis predicts no shock formation \eqref{BernoulliBlowup}. The slow scale analysis \eqref{BurgersDistSin} decently predicts shock formation at the time $(t^*)^\text{pol} \approx 0.077$~s obtained by numerical optimisation, and the distance $(Y^*)^\text{pol} \approx 0.16$~m (critical amplitude $56.03 < |A|$). Numerical results suggest also that accelerations decay over time for the smallest amplitude, which is coherent with the slow scale theory (critical amplitude $39.63 > |A|$). If we refer to the analysis of the elastic limit (Appendix~\ref{app:Elastic}), the present simulations correspond to the loading Mach numbers $\text{Ma} \approx 0.92$ and $\text{Ma} \approx 0.60$, which belong to a range of parameters where the slow scale theory is still fairly accurate in the absence of dissipation.

In the the low-dissipation limit, the accuracy of the slow scale theory was already benchmarked in the Appendix~\ref{app:Elastic}. In the viscoelastic case, this analytical estimate should be used with caution too, as its apparent accuracy does not extend to the full range of loading signals. More specifically, the ramp signal case $\Omega\to 0$ highlights the limitations of the slow scale theory at low frequency (Figs.~\ref{fig:RampShockElast}-\ref{fig:RampShock}). Similar observations can be made for the exponential model, leading to the final assessment in Table~\ref{tab:Summary}.

\section{Conclusion}\label{sec:Conclu}

For incompressible viscoelastic solids, analytical shock formation conditions have been derived, and a finite volume method has been implemented. Numerical simulations with ramp signal show that both the acceleration wave analysis and the slow scale analysis provide inaccurate shock time and shock distance estimates. Shock formation could occur even if acceleration waves do not blow up, as was already found in the non-dissipative case \cite{jordan05} (cf. Table~\ref{tab:Summary}).

The growth and decay of discontinuities might be further investigated using dedicated techniques \cite{schuler73,fu90}. Future works could also encompass the study of travelling waves \cite{depascalis19}, steady shocks \cite{schuler73}, and solitary waves in layered media \cite{ziv20}. Results could be extended to pre-strained solids \cite{pucci19}, other viscoelastic material models \cite{derooij16}, poroelasticity \cite{berjamin21b}, and poro-viscoelasticity \cite{comellas20,hosseini20}. Lastly, higher-order and adaptive computational methods could be used for increased performance in multi-dimensional configurations \cite{tripathi19, reinarz20}.

\subsection*{Acknowledgments}

The authors are grateful to Tal Cohen (MIT) and Michel Destrade (NUI Galway) for support. HB's work was supported by the Irish Research Council [project ID GOIPD/2019/328].



\bibliographystyle{unsrtnat}
\bibliography{biblio}

\appendix

\section{Linear solution}\label{app:Linear}

Linearising Eq.~\eqref{SystHypVect} about an undeformed equilibrium state, the first-order system \eqref{SystHypVect} becomes ${\bf q}_t + {\bf A}\, {\bf q}_Y = {\bf B}\,  {\bf q}$ with
\begin{equation}
	{\bf q} = \begin{bmatrix}
		\gamma\\
		v\\
		r\end{bmatrix} ,\qquad
	{\bf A} = \begin{bmatrix}
		0 & -1 & 0 \\
		-{\mu}/{\rho} & 0 & 1/\rho\\
		0 & 0 & 0
	\end{bmatrix} , \qquad
	{\bf B} = \frac{1}{\tau} \begin{bmatrix}
		0 & 0 & 0 \\
		0 & 0 & 0 \\
		g \mu & 0 & -1
	\end{bmatrix} .
\end{equation}
Fourier transformation of the above PDE leads to the dispersion relationship
\begin{equation}
	\rho\frac{\omega^2}{\kappa^2} = \mu\left( 1 - \frac{g}{1+\text{i}\omega\tau}\right) ,
	\label{Dispersion}
\end{equation}
where $\omega$ denotes the angular frequency, $\kappa$ is the wavenumber, and $\text{i} = \sqrt{-1}$ is the imaginary unit. Here, we have defined Fourier transformation in time as $\int (\cdot) \text{e}^{-\text i\omega t}\text dt$ (for which the hat symbol is used later on), and spatial Fourier transformation as $\int (\cdot) \text{e}^{\text i\kappa Y}\text dY$.

Dispersion and dissipation properties can be deduced from the dispersion relationship $\rho {\omega^2}/{\kappa^2} = M$ of Eq.~\eqref{Dispersion}, see Carcione \cite{carcione15}. Using the expression of the dynamic modulus $M$, one derives the \emph{dissipation factor}
\begin{equation}
	D = \frac{\text{Im}\, M}{\text{Re}\, M} = \frac{g \omega \tau}{1 - g + (\omega \tau)^2} \, .
\end{equation}
The maximum value $\Omega_1/\Omega^\text{v}$ of the dissipation factor is reached at the characteristic frequency $\omega = \Omega^\text{v}$ given by $\Omega^\text{v} = \sqrt{1-g}/\tau$, and $\Omega_1 = g/(2\tau)$ is the decay rate of viscoelastic acceleration waves \eqref{BernoulliCoeff}.

Partial Fourier transformation of the equations of motion in time domain yields the harmonic oscillator equation $\partial^2_{Y} \hat v + \kappa^2 \hat v = 0$ for the velocity field, where the wavenumber $\kappa$ is a root of the quadratic equation \eqref{Dispersion} to be specified. Integration of the differential equation for $\hat v$ with $\hat v|_{Y=0} = \hat V$ gives $\hat v = \hat V\, \text{e}^{-\text{i}\kappa Y} + \hat\vartheta\sin(\kappa Y)$ where $\hat\vartheta$ is arbitrary. To make the contribution of $\hat V$ decay as $Y$ increases towards infinity, the wavenumber $\kappa$ is necessarily equal to the principal square root of $\kappa^2$. Then, the decay of $\hat v$ at infinity requires $\hat\vartheta \equiv 0$, that is $\hat v = \hat V\, \text{e}^{-\text{i}\kappa Y}$. Inverse Fourier transformation in time yields the following integral representation of the solution
\begin{equation}
    v(Y,t) = \frac{1}{2\pi} \int_{\mathbb R} \hat V(\omega)\, \text{e}^{\text i (\omega t - \kappa Y)} \text d\omega = \frac{1}{\pi} \int_{0}^{\infty} \text{Re}\left( \hat V(\omega)\, \text{e}^{\text i (\omega t - \kappa Y)} \right) \text d\omega ,
	\label{SolLin}
\end{equation}
where the Hermitian symmetry of $\hat v$ was used (the velocity $v$ is real). Of course, the linear elastic solution $V(t-Y/c_0)$ is recovered in the limits $g=0$, $\tau\to +\infty$, where the wavenumber reduces to $\kappa = \omega/c_0$ with $c_0 = \sqrt{\mu/\rho}$ denoting the shear wave speed. In practice, the above integrals \eqref{SolLin} are evaluated using a Fast Fourier Transform algorithm.

\section{Elastic limit}\label{app:Elastic}

Let us recall the main properties regarding the formation of shocks in the elastic case for which the memory variables $r$, $s$ vanish (i.e., $g =0$ or $\tau \to + \infty$). Thus, the shear stress component $\sigma$ is function of $\gamma$ only. Injecting the Ansatz $v = \phi(\gamma)$ in the first two lines of Eq.~\eqref{SystHypVect} yields the condition \cite{chu64}
\begin{equation}
	v \mp Q = R^\pm , \qquad
	Q = \int_0^\gamma c(\vartheta)\, \text d\vartheta , \qquad
	c(\gamma) = \sqrt\frac{\sigma'(\gamma)}{\rho} ,
	\label{InvarElast}
\end{equation}
where $R^\pm = \phi(0)$ is an arbitrary constant, and primes denote differentiation with respect to the argument. In terms of the Riemann invariants $R^\pm$, the quasi-linear system governing elastic wave propagation is diagonal, with scalar components $\partial_t R^\pm \pm c(\gamma)\, \partial_Y R^\pm = 0$.

The `$\pm$' characteristic curve starting at $Y=0$ and $t=t_0^\pm$ carries a constant value of $R^\pm = f^\pm(t_0^\pm)$, where $f^\pm$ is an arbitrary function. The impinging wavefront corresponds to the curve $t_0^+ = 0$ along which $v$, $Q$ equal zero according to the boundary condition \eqref{BCond}. In particular, $R^\pm$ equals zero along this curve, and we can conclude that $R^-$ is uniformly zero. Since $t_0^\pm =t_0$ yields $v = V(t_0)$ at the boundary, we know also that $Q = -V(t_0)$. Therefore, the quantities $v$, $Q$ as well as $\gamma$, $c$ may be viewed as functions of $t_0$, and the `$+$' characteristics are straight lines with equation $Y = c(t_0) (t-t_0)$.

Differentiation with respect to $t_0$ then provides the condition $c'(t_0^*) > 0$ for the formation of shocks at positive time $t^* = t^\dagger(t_0^*)$ and position $Y^* = Y^\dagger(t_0^*)$ given by \cite{chockalingam20}
\begin{equation}
	t_0^* = \underset{t_0 > 0}{\arg\inf}\; t^\dagger(t_0) = \underset{t_0 > 0}{\arg\inf}\; Y^\dagger(t_0) ,
	\qquad\text{where}\qquad
	t^\dagger(t_0) = t_0 + \frac{c(t_0)}{c'(t_0)} >0, \qquad
	Y^\dagger(t_0) = \frac{c(t_0)^2}{c'(t_0)} >0 ,
	\label{ShockChar}
\end{equation}
and $c' = \gamma'\, ({\text d c}/{\text d\gamma})$. Since the strains $\gamma$ are linked to the input signal \eqref{BCond} according to $Q = -V(t_0)$ along characteristics, differentiation with respect to $t_0$ on both sides yields $\gamma' = -V'/c$, and Eq.~\eqref{ShockChar} is modified according to
\begin{equation}
	t^\dagger(t_0) = t_0 - \frac{c(t_0)^2}{V'(t_0) \frac{\text d c}{\text d\gamma}(t_0)} >0, \qquad
	Y^\dagger(t_0) = \frac{-c(t_0)^3}{V'(t_0) \frac{\text d c}{\text d\gamma}(t_0)} >0 .
	\label{ShockElast}
\end{equation}
Hence, shock formation combines the loading programme (through $V'$) and the material behaviour (through ${\text d c}/{\text d\gamma}$).

Using the exponential strain energy function $W^\text{exp}$ in Eq.~\eqref{W}, one obtains \cite{chockalingam20}
\begin{equation}
	\sigma(\gamma) = \frac{\mu}{\alpha} \left(\text{e}^{\alpha |\gamma|} - 1\right) \operatorname{sign}(\gamma)
	\qquad\text{and}\qquad
	c = c_0\, \text{e}^{\frac\alpha{2} |\gamma|} , \quad
	\frac{\text d c}{\text d \gamma} = \frac\alpha{2} c \operatorname{sign}(\gamma) , \quad
	Q = \frac{c-c_0}{\alpha/2} \operatorname{sign}(\gamma) ,
	\label{ExpCalc}
\end{equation}
where $c_0 = \sqrt{\mu/\rho}$ is the speed of linear shear waves. Here, the inverse function of $\gamma \mapsto Q(\gamma)$ can be expressed analytically, and we have $\gamma = \frac2{\alpha} \ln \big( 1 + \frac{\alpha}{2c_0} |Q| \big) \operatorname{sign}(Q)$.
With the polynomial model $W^\text{pol}$, we find
\begin{equation}
	\sigma(\gamma) = \mu\left(1 + \tfrac13 (b \gamma)^2\right) \gamma
	\qquad\text{and}\qquad
	c = c_0 \sqrt{1 + (b\gamma)^2}\; , \quad
	\frac{\text d c}{\text d \gamma} = \frac{(b c_0)^2 \gamma}{c} , \quad
	Q = \frac{\gamma}{2}  \left(c + c_0\frac{\operatorname{arsinh} (b\gamma)}{b\gamma}\right),
	\label{PolCalc}
\end{equation}
where $b = \sqrt{6 \beta C_1/\mu}$. No analytical expression for the inverse function of $\gamma \mapsto Q(\gamma)$ is known in this case.

Now, if we specialise Eq.~\eqref{ShockElast} for the exponential model \eqref{ExpCalc} with ramp-type or sinusoidal loading \eqref{Sign}, we find the breaking time and shock distance
\begin{equation}
	(t^*)^\text{exp} = \frac{2c_0}{\alpha \left|A\right|} > 0 ,
	\qquad 
	(Y^*)^\text{exp} = c_0\, (t^*)^\text{exp} > 0 ,
	\label{ShockElastExp}
\end{equation}
which are reached along the first characteristic line $t_0^* = 0$.
Given that computations are quite straightforward with the exponential model, such results can be extended to other loading programmes, as well as to the localisation of finite acceleration levels \cite{chockalingam20}. Unfortunately, a similar derivation is not tractable analytically for the polynomial model \eqref{PolCalc} because the inverse mapping of $\gamma \mapsto Q(\gamma)$ is transcendental.

If shock formation shall occur at the edge of the impinging signal ($t_0^* = 0$), then Eq.~\eqref{ShockElast} yields
\begin{equation}
	t^* = -\frac{c_0^2}{A\, \frac{\text d c}{\text d\gamma} \big|_{\gamma=0}}  > 0 ,
	\qquad 
	Y^* = c_0 t^* > 0
	\label{ShockElastAcc}
\end{equation}
for the ramp and sinusoidal signals \eqref{Sign}. Thus, the condition for shock formation at the wavefront reduces to the requirement for $A$ and $\frac{\text d c}{\text d\gamma} \big|_{\gamma=0}$ to have opposite signs. Using the asymptotic equivalence $\gamma \sim Q/c_0$ in the vicinity of $\gamma = 0$, we conclude that the exponential model \eqref{ExpCalc} entails shock formation within a finite range \eqref{ShockElastExp} along $t_0^* = 0$, while the polynomial model \eqref{PolCalc} prevents shock formation along $t_0^* = 0$, in general.

As shown by the above computations, if a shock forms within a finite distance of propagation inside a polynomial solid, then it must happen beyond the wavefront's edge.
In this case ($t_0^* > 0$), the optimality condition \eqref{ShockElast} yields
\begin{equation}
	\frac{\text d^2 c}{\text d \gamma^2} - \frac{3}{c} \left(\frac{\text d c}{\text d \gamma}\right)^2 = \frac{V''(t_0^*)}{V'(t_0^*)^2}\, c \frac{\text d c}{\text d \gamma}
	\label{Opt}
\end{equation}
upon differentiation with respect to $t_0$, where we have used the relationship $\gamma' = -V'/c$. The expressions in Eq.~\eqref{PolCalc} then provide the shear strain $\gamma^* = \gamma(t_0^*)$ which solves the above equation.

For the ramp signal $\Omega\to 0$, Eq.~\eqref{Opt} simplifies greatly due to $V'(t_0^*) = A$ and $V''(t_0^*) = 0$. With the polynomial model, the solutions $\gamma^* = \pm 1/(b\sqrt{3})$ are found. Using $Q = -At_0^*$, Eq.~\eqref{PolCalc} gives $t_0^* = \big(\frac13 + \frac12 \operatorname{arsinh} \frac{1}{\sqrt{3}} \big) \frac{c_0}{b \left|A\right|}$, and thus
\begin{equation}
	(t^*)^\text{pol} = t_0^* + \frac83 \frac{c_0}{b \left|A\right|} \simeq 3.27\, \frac{c_0}{b \left| A\right|} > 0, \qquad
	(Y^*)^\text{pol} = \frac{16}{3\sqrt{3}} \frac{c_0^2}{b \left|A\right|} \simeq 3.08\, \frac{c_0^2}{b \left|A\right|} > 0 .
	\label{ChockaShockRamp}
\end{equation}
Note that such a shock is necessarily located behind the wavefront's edge (i.e., $c_0\, (t^*)^\text{pol} > (Y^*)^\text{pol}$).

For the sinusoidal signal $\Omega > 0$, these computations are more involved. Following the approach in \cite{chockalingam20}, we rescale Eq.~\eqref{Opt} and $Q = -V(t_0)$ by introducing the dimensionless quantities $\bar c = c/c_0$, $\bar V = V \Omega/A$, $\bar t = \Omega t$, $\bar Q = b Q/c_0$ and $\delta = b \gamma$. Thus, the following expression for the dimensionless shock time $\bar t^* = \Omega t^*$ and distance $\bar Y^* = \Omega Y^*/c_0$ can be derived
\begin{equation}
	(\bar t^*)^\text{pol} = -\arcsin\left(\frac{\bar Q^*}{\text{Ma}}\right) + \frac{(\bar Y^*)^\text{pol}}{\sqrt{1+(\delta^*)^2}} , \qquad
   (\bar Y^*)^\text{pol} = -\frac{\big(1+(\delta^*)^2\big)^2}{\delta^* \operatorname{sign}(\text{Ma}) \sqrt{\text{Ma}^2 - (\bar Q^*)^2}} ,
   \label{ChockaShock}
\end{equation}
where $\text{Ma} = \frac{b A}{c_0 \Omega}$ is a loading Mach number, $\bar Q^* = \bar Q|_{\delta=\delta^*}$, and $\delta^*$ is a suitable root of
\begin{equation}
    3 - \frac{1}{\delta^2} + \frac{(1+\delta^2)^{3/2}}{\delta} \frac{\bar Q}{\text{Ma}^2 - \bar Q^2} = 0
    \qquad\text{with}\qquad
    \bar Q = \frac{1}{2} \left(\delta \sqrt{1+\delta^2} + \operatorname{arsinh}(\delta) \right) .
\end{equation}
Consistently with the ramp signal case $\Omega \to 0$ \eqref{ChockaShockRamp}, the scaled strain $\delta^*$ has the asymptotic value $\pm 1/\sqrt{3}$ as $|\text{Ma}| \to +\infty$.

\begin{figure}
    \begin{minipage}{0.49\linewidth}
        \centering
        (a)
        
        \includegraphics{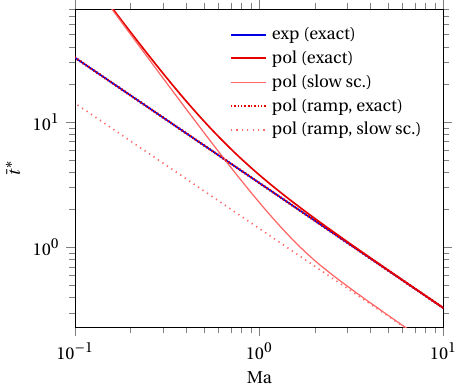}
    \end{minipage}\hfill
    \begin{minipage}{0.5\linewidth}
        \centering
        (b)
        
        \includegraphics{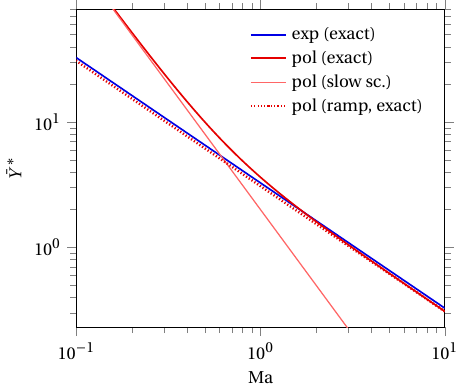}
    \end{minipage}
    
    \caption{Shock formation analysis in the elastic limit. Evolution of dimensionless shock time (a) and shock distance estimates (b) in terms of the loading Mach number.}
    \label{fig:ChockaShock}
\end{figure}

For polynomial elastic solids, the dimensionless shock time and distance \eqref{ChockaShock} is plotted as a function of the loading Mach number $\text{Ma}$ in Figure~\ref{fig:ChockaShock}. Obviously, the ramp signal case \eqref{ChockaShockRamp} provides the correct asymptotes $(\bar t^*)^\text{pol} \sim 3.27/\text{Ma}$ and $(\bar Y^*)^\text{pol} \sim 3.08/\text{Ma}$ at large Mach numbers (densely dotted lines). By design, the exponential model with the connection $\alpha \approx 0.612\, b$ yields the same shock times $(\bar t^*)^\text{exp} = 3.27/\text{Ma}$, but slightly larger shock distances $(\bar Y^*)^\text{exp} = 3.27/\text{Ma}$ matching the position of the wavefront \eqref{ShockElastExp}. If we compare \eqref{ChockaShock} with the elastic slow scale estimates \eqref{BurgersDistSin} expressed as $(\bar Y^*)^\text{pol} = 2/\text{Ma}^2$ and $(\bar t^*)^\text{pol} = (\bar Y^*)^\text{pol}/\sin(2\bar t^*_0) + \bar t^*_0 - \frac12 \tan(\bar t^*_0)$ with $\bar t^*_0 = \min \big\lbrace \frac{\pi}4, \arcsin\big(\sqrt{2}/|\text{Ma}|\big) \big\rbrace$, then we find that the latter are accurate at low Mach numbers only (thin light curves). At large Mach numbers, e.g. for ramp signal $\Omega\to 0$ where Eq.~\eqref{BurgersDistRamp} applies, the slow scale analysis underestimates shock times and distances.

\end{document}